\documentclass[%
 reprint,
superscriptaddress,
showpacs,preprintnumbers,
 amsmath,amssymb,
 aps,
pra,
]{revtex4-1}
\usepackage[dvipsnames]{xcolor}
\usepackage[colorlinks=true,bookmarks=false,linkcolor=NavyBlue,urlcolor=NavyBlue,citecolor=NavyBlue,breaklinks]{hyperref}
\usepackage{amsmath}
\usepackage{amsfonts}
\usepackage{graphicx}
\usepackage{color}
\usepackage{braket}
\usepackage{graphicx}
\usepackage{dcolumn}
\usepackage{bm}

\begin{document}

\title{Effect of surface state hybridization on current-induced spin-orbit torque in thin topological insulator films}
\author{Cong Son Ho}
 \email{elehcs@nus.edu.sg}
 \affiliation{
Department of Electrical and Computer Engineering, National University of
Singapore, 4 Engineering Drive 3, Singapore 117576.}

\author{Yi Wang}
 \affiliation{
Department of Electrical and Computer Engineering, National University of
Singapore, 4 Engineering Drive 3, Singapore 117576.}

\author{Zhuo Bin Siu}%
\affiliation{
Department of Electrical and Computer Engineering, National University of
Singapore, 4 Engineering Drive 3, Singapore 117576.}

\author{Hyunsoo Yang}%
\affiliation{
Department of Electrical and Computer Engineering, National University of
Singapore, 4 Engineering Drive 3, Singapore 117576.}

\author{Seng Ghee Tan}
\affiliation{
Data Storage Institute, Agency for Science, Technology and Research (A*STAR), 2
Fusionopolis Way, 08-01 Innovis, Singapore}
\author{Mansoor B. A. Jalil}%
 \email{elembaj@nus.edu.sg}
\affiliation{
Department of Electrical and Computer Engineering, National University of
Singapore, 4 Engineering Drive 3, Singapore 117576.}

\date{\today}

\begin{abstract}
We investigate the current-induced spin-orbit torque in thin topological insulator (TI) films in
the presence of hybridization between the top and bottom surface states. We
formulate the relation between spin torque and TI thickness, from which we
derived the optimal value of the thickness to maximize the torque. We show
numerically that in typical TI thin films made of $\mathrm{Bi_2Se_3}$,
the optimal thickness is about 3-5 nm.
\end{abstract}

\maketitle


\section{Introduction}

Spin torque is one of the most actively researched topics in spintronics. In
spin-orbit coupling (SOC) systems, the resulting spin-orbit torque (SOT) comprises of
two types: field-like torque \cite{Tan:arxiv07,*Tan:ann11,Manchon:prb08,Matos:prb09, Miron:nat10,Li:prb15} and
damping-like torque \cite{Haney:prb13,Kure:nat14}. The SOT strength scales with the SOC
strength, therefore one can obtain a large SOT signal in strong SOC systems, such
as heavy metal/ferromagnet (Pt/Co) \cite{Mihai:nat10} and topological insulators (TI)
\cite{Mell:nat14,Wang:prl15,Fan:nat14}.

Topological insulators possess strong SOC which translates to a large SOT
effect. Moreover, the spin-locked topological surface states of TI are robust
under effects of the time-reversal symmetric impurity scattering. These factors
enable TI to be one of the most promising candidates for SOT devices. When a
current is passed onto the TI surface (interface), a non-equilibrium spin
density will be induced on that surface \cite{Mell:nat14}, which is directed 
in-plane and perpendicular to the applied current as a
result of the spin-momentum locking of the surface states \cite{Mell:nat14}, 
similar to the Rashba-Edelstein effect \cite{Edel:ssc90}. Thus spin
 accumulation is proportional to the charge current flowing on the
 TI surface \cite{Tan:arxiv07,Manchon:prb08,Mell:nat14}. In addition,
 an out-of-plane spin density can be generated due to the
Berry curvature of the spin texture in the momentum space \cite{Kure:nat14,Mura:sci03,Fujita:njp10},
 and  due to the presence of the hexagonal-warping effect \cite{Kuro:prl10}.  The two components, in-plane and out-of-plane spin density,
are responsible for inducing an out-of-plane and in-plane torques, respectively.

In the study of current-induced SOT, spin torque efficiency (ratio of torque
strength and current density in the TI - $\mathcal{T}/j_\mathrm{TI}$, which 
is equivalent to spin Hall angle $j_s/j_\mathrm{TI}$) is a crucial quantity.
 In 3D TI's, the surface state has a finite thickness and usually
extends into the bulk material, up to 1 nm from the TI-vacuum interface
\cite{Wang:prl15,Lin:prb09}, and the current is supposed to flow entirely in the surface channel in ideal TI. However, in
practical 3D TI's, the bulk is not totally insulating \cite{Ban:prl12}, 
and thus the current in the TI can flow either in the surface and bulk channels. Since the bulk states do not exhibit any
spin-momentum locking, TI with a large bulk conductance will induce a smaller torque on an
adjacent ferromagnetic (FM) layer. Therefore, 
it is obvious that one has to reduce the TI thickness
to reduce the relative contribution of the bulk
\cite{Zhang:apl09,Peng:nat10,Zhang:nat10}, and enhance the torque efficiency.
But how far can the thickness of a TI film be reduced?

Normally, a 3D TI film comprises of two surfaces (top and bottom surfaces, for
example). In a typically thick TI film, the two surfaces are uncorrelated, i.e,
the transport on one surface will not affect the other. However, if the film is
made very thin, {\it{i.e.}}, its thickness is comparable to the decay length of
surface states, the two surface states become hybridized \cite{Linder:prb09,Lu:prb10,Liu:prb10, Zyu:prb11}. This
hybridization is quantified by a tunneling coupling constant ($\mathrm{\Delta
}$), which is thus a function of the TI thickness. Generally,
$\mathrm{\Delta }$ is larger as the thickness is reduced \cite{Linder:prb09,Lu:prb10,Liu:prb10}. The
effect of the surfaces hybridization on spin torque is as follows. When a charge
current is passed onto the top surface in the $x$-direction, it can leak to the
bottom surface through the tunneling effect \cite{Persho:prb12}. While the
current on the top surface induces a spin accumulation in the $y$-direction, its
bottom counterpart induces a spin accumulation in the opposite, {\it{i.e.}},
$-y$-direction, since the two surface states have opposite helicities. As a
result, the total spin accumulation is reduced with the hybridization effect.
Thus, the spin torque is also reduced as the TI thickness is
reduced. This dependence will be formulated subsequently in this paper.

From the above, we may surmise that the reduction in the TI thickness has two
competing effects on the spin torque: (i) the torque strength is stronger as the
bulk contribution is decreased; (ii) on the other hand, the torque strength
becomes suppressed due to increasing hybridization of the surface states.
Therefore, there is an optimal value of TI thickness which would maximize the
torque efficiency.

Thus, the aim of this study is to investigate the spin torque in TI film in the
presence of the hybridization effect. We will formulate the relation between
spin torque and TI thickness, from which we derive the optimal value of the
thickness to maximize the torque. For the exemplary cases of
$\mathrm{Bi_2Se_3}$ or $\mathrm{Bi_2Te_3}$ thin films, the optimal thickness is found to be
about 3-5 nm. In previous experiments \cite{Mell:nat14,Wang:prl15,Fan:nat14},
the TI films have thickness in the range of 6-20 nm, which is far above the
predicted optimal values. The paper is organized as follows: in Section
\ref{theory}, we formulate the theory of spin torque in a topological insulator
(TI) system coupled to a ferromagnetic (FM) layer. In Section \ref{hybri}, we
will apply the theory to the thin TI-FM film, where the hybridization between
the top and bottom surface states is taken into consideration, and derive the
optimal thickness for maximum torque efficiency. The conclusion is given in the
last section.

\section{Theory}\label{theory}
We will first formulate the spin-torque induced by charge current in a coupled
FM/TI bilayer system in the absence of inter-surface hybridization. The
Hamiltonian of the system is given by:
\begin{equation}\label{TI1}
\mathcal{H}={\mathcal{H}}_\mathrm{TI}+J_\mathrm{ex}\mathbf{m}\cdot \hat{\sigma
},
\end{equation}
where the first term is the Dirac Hamiltonian of the topological surface state
${\mathcal{H}}_\mathrm{TI}=\hbar v_\mathrm{F}\left(\mathbf{k}\times
\hat{z}\right)\cdot \hat{\sigma }$\textbf{ } with $v_\mathrm{F}$ being the Fermi
velocity, $\hat{\sigma }$ is the vector the Pauli matrices. The last term is the
$s$-$d$ exchange interaction between itinerant electron and magnetization with
$J_\mathrm{ex}$ being the exchange constant. We can rewrite Eq. \eqref{TI1} to
express electron spin in an effective field as $\mathcal{H}=\mathbf{b}\cdot
\hat{\sigma }$, where $\mathbf{b}=\hbar v_\mathrm{F}\left(\mathbf{k}\times
\hat{z}\right)+J_\mathrm{ex}\mathbf{m}$. The eigenenergies are found as
\begin{equation}\label{TI2}
{\mathcal{E}}_s\left(\mathbf{k}\right)=s\sqrt{{\left(\hbar
v_\mathrm{F}k\right)}^2+{\left(J_\mathrm{ex}\right)}^2+2\hbar
v_\mathrm{F}J_\mathrm{ex}\mathbf{m}\cdot \left(\mathbf{k}\times \hat{z}\right)}
\end{equation}
where $s=\pm 1$ which is for majority and minority electron, respectively, and
we have used the fact that $|{\mathbf m}|=1$. For a given value of momentum
$\mathbf{k}$, Eq. \eqref{TI2} can be considered as the energy of the magnetic
system. Thus, the torque field acting on the magnetization can be found by
taking the functional derivative of $\mathcal{E}_s(\mathbf{k})$, {\it{i.e.}}
\cite{Mansoor16},
\begin{equation}\label{TI3}
{\mathbf{H}}^\mathrm{eff}=-\frac{n}{{\mu }_0M_s}\frac{\delta \mathcal{E}}{\delta
\mathbf{m}}.
\end{equation}
In the above, $n$ is the charge density, and $M_s$ is the saturated
magnetization. Let us first consider the contribution from the lower band. In this
case, from Eqs. \eqref{TI2} and \eqref{TI3}, the torque field is
\begin{equation} \label{GrindEQ__4_} 
{\mathbf{H}}^\mathrm{eff}=\frac{n\hbar v_\mathrm{F}}{{\mu
}_0M_s}\left(\mathbf{k}\times \hat{z}\right).
\end{equation} 
The upper band will gives rise to an opposite torque field. To obtain the above, we have assumed the strong exchange limit, i.e,
$J_\mathrm{ex}\gg \hbar v_\mathrm{F}k_\mathrm{F}$, $|b|\approx J_\mathrm{ex}$, and the
effective field is taken to the first order in $\hbar
v_\mathrm{F}k/J_\mathrm{ex}$. The net torque field is found by summing over all
momentum space. In the absence of applied current, the expectation value
of\textbf{ }$\mathbf{k}$ would vanish and so would the effective field. However,
in the presence of some charge current ${\mathbf{j}}^{ss}_e$ that flows on the
surface, the average $\mathbf{k}$ is finite and can be found as following. The
charge current is expressed as ${\mathbf{j}}^{ss}_e=ne\left\langle
\mathbf{v}\right\rangle $, where $n$ is the sheet carrier density, and
$\mathbf{v}=\frac{\partial H}{\hbar \partial k}=v_\mathrm{F}\left(\hat{z}\times
\hat{\sigma }\right)$ is the velocity. In the adiabatic limit, the electron spin
is mostly aligned along the effective field as $\left\langle \hat{\sigma
}\right\rangle =\mathbf{b}/|b|$, from which the charge current is
${\mathbf{j}}^{ss}_e=ne\frac{\hbar v^2_F}{J_\mathrm{ex}}\left\langle
\mathbf{k}\right\rangle +nev_\mathrm{F}\left(\hat{z}\times \mathbf{m}\right)$.
Rearranging the equation, we have
\begin{equation} \label{GrindEQ__5_} 
\left\langle \mathbf{k}\right\rangle =\frac{J_\mathrm{ex}}{ne\hbar
v^2_F}{\mathbf{j}}^{ss}_e-\frac{J_\mathrm{ex}}{\hbar
v_\mathrm{F}}\left(\hat{z}\times \mathbf{m}\right).
\end{equation} 
Substituting Eq. \eqref{GrindEQ__5_} into Eq. \eqref{GrindEQ__4_}, we obtain the
current-induced torque field,
\begin{equation} \label{GrindEQ__6_} 
{\mathbf{H}}^\mathrm{eff}=\eta_0\left({\mathbf{j}}^{ss}_e\times \hat{z}\right),
\end{equation} 
where $\eta_0=\frac{J_\mathrm{ex}}{{ev_\mathrm{F}\mu }_0M_s}$, and we have ignored 
the component that is parallel to $\mathbf{m}$, as this
does not induce torque on the magnetization itself. The field in Eq.
\eqref{GrindEQ__6_} is in-plane and it induces an out-of-plane torque on the
magnetization as $\mathcal{T}=\mathbf{m}\times {\mathbf{H}}^\mathrm{eff}$.
Explicitly,
\begin{equation} \label{GrindEQ__7_} 
\mathcal{T}=\eta_0\mathbf{m}\times \left({\mathbf{j}}^{ss}_e\times
\hat{z}\right).
\end{equation}

\begin{figure}
	\includegraphics[width=0.5\textwidth]{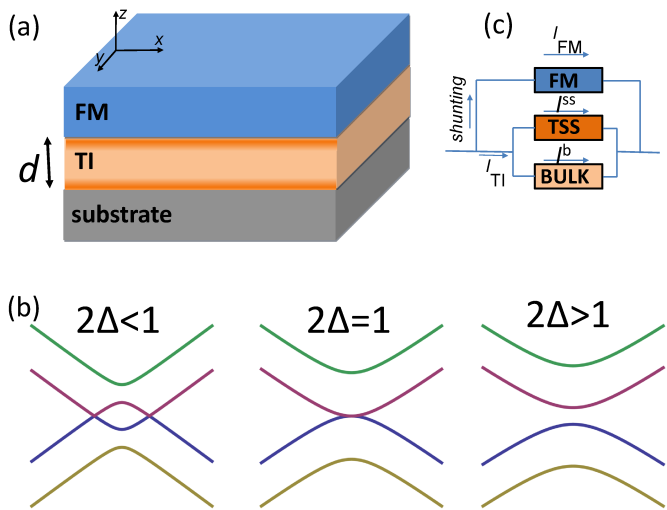}
\caption{(Color online) (a) Schematic diagram of bilayer ferromagnet/topological
insulator. The TI film is of the thickness $d$. (b) Schematic energy $\mathcal E$ versus
$k_x$ subbands near the Dirac points ($k_x=0$) for various values of the
tunneling element $\Delta$. (c) Schematic diagram of the current distribution in the
 bilayer with $I_{TI}$ being the current in the TI film, and $I_{FM}$ being the shunting 
current through the FM layer. In the TI film, the current can flow in the surface state 
channel $(I^{ss})$ and bulk channel $(I^{b})$, respectively.\label{Fig1}}
\end{figure}

In practice, an applied current $j^0_e$ in the system is composed of three
different channels: (i) surface current $j^{ss}_e$, (ii) bulk current $j^b_e$,
and (iii) shunting current through the FM layer $j^{\mathrm{FM}}_e$, see Fig. \ref{Fig1}. Therefore,
out of the total applied charge current $j^0_e$, only $j^{ss}_e$ would generate
a torque field and spin torque, following Eqs. \eqref{GrindEQ__6_} and
\eqref{GrindEQ__7_}. To obtain the relation between the surface current and the
net current, we assume that the FM and TI layers have thicknesses $t$ and $ d$,
and conductances $G_{\mathrm{FM}}$ and $ G_\mathrm{TI}$, respectively. The
surface current is then given by
\begin{equation} \label{GrindEQ__8_} 
j^{ss}_e\left(d\right)=j^0_e\frac{G^{ss}_\mathrm{TI}}{G_\mathrm{TI}+G_{\mathrm{FM}}}=j_{TI}\frac{G^{ss}_\mathrm{TI}}{G_\mathrm{TI}}.
\end{equation} 
In the above, $j_{TI}$ is the current flowing in the TI film, $\
G^{ss}_\mathrm{TI}$ is the conductance of TI surface, which is assumed to be
unchanged as the thickness of the TI layer $d$ is changed. Meanwhile, the total
conductance of the TI layer $G_\mathrm{TI}$ increases with $d$ \cite{Ban:prl12},
reflecting the increased contribution of the bulk channel.
Therefore, the spin torque and torque field would generally be enhanced as $d$
is reduced. In practice, TI films can be made very thin, {\it{i.e.}}, down to 2
nm \cite{Ban:prl12}, and therefore one may induce large spin torque in such
systems. However, as the TI film thickness is reduced to below some critical
value, the property of the surface state will be changed, {\it{i.e.}}, an energy
gap is opened and the Dirac cone would disappear, as a result of the surface
hybridization \cite{Lin:prb09,Zhang:nat10,Zyu:prb11,Lu:prb10}. As a consequence,
all spin transport effects arising from the topological surface states will be
modified accordingly. The effect of surface hybridization will be considered in
the following section.

\section{SPIN TORQUE WITH HYBRIDIZATION EFFECT}\label{hybri}
\subsection{Effect of surface hybridization}

In this Section, we will formulate the spin-orbit torque in a thin TI film with
hybridization effect. In this case, both surface states (of top and bottom
surfaces) are taken into account. We assume that the magnetic layer is only in
contact with the top TI surface with the exchange coupling $J_\mathrm{ex}$. The
itinerant electron on the bottom surface may also couple to the FM, but with a
weaker exchange coupling, which is then ignored for simplicity. In addition,
we also disregard the asymmetry between the two interfaces, {\it i.e.}, FM/TI
and TI/substrate interfaces.

The Hamiltonian of the system is then given by  \cite{Zyu:prb11}
\begin{equation} \label{GrindEQ__9_} 
\mathcal{H}=\left[ \begin{array}{cc}
{\mathcal{H}}_\mathrm{TI}+J_\mathrm{ex}\mathbf{m}\mathbf{\cdot }\hat{\sigma } &
\mathrm{\Delta }I_2 \\
\mathrm{\Delta }I_2 & -{\mathcal{H}}_\mathrm{TI}\end{array}
\right]. 
\end{equation} 
In the above, $\mathrm{\Delta }$ represents the tunneling element
(hybridization) between top and bottom surfaces, $I_2$ is the $(2\times 2)$ unit
matrix. The value of $\mathrm{\Delta }$ has an inverse relation with the
thickness of the TI film, e.g, $\Delta=0.05\ \mathrm{eV}$ for $d=5$ nm film, but
increases to 0.25 eV for $d=2$ nm \cite{Peng:nat10}. In the range of small
thickness, the tunneling element can be approximated as \cite{Linder:prb09,Lu:prb10}
\begin{equation} \label{GrindEQ__10_} 
\mathrm{\Delta }\approx\frac{B_1{\pi }^2}{d^2}, 
\end{equation} 
where $B_1$ is a material-dependent parameter \cite{Zhang:nat10}. Note that the
Fermi velocity is also thickness-dependent. However, the variation can be
neglected as its values in the thick and thin film limits only differ by the
order of 0.01 \cite{Lu:prb10}.

From the Hamiltonian in Eq. \ref{GrindEQ__9_} , the eigenenergies are given by
\begin{equation} \label{GrindEQ__11_} 
{\mathcal{E}}_{s\tau }=s\sqrt{U+\tau V}, 
\end{equation} 
with $s,\tau =\pm 1$, and 
\begin{eqnarray}
U&=&{\mathrm{\Delta }}^2+\frac{J^2_\mathrm{ex}}{2}+{\hbar
}^2{v^2_F\mathbf{{k}}}^2+\hbar
v_\mathrm{F}J_\mathrm{ex}{\left(\mathbf{{m}}\times\mathbf{{k}}\right)}_z,\nonumber\\
V&=&\frac{1}{2}\sqrt{\mathrm{4}{\mathrm{\Delta
}}^{\mathrm{2}}\mathrm{+}{\left(J_\mathrm{ex}+2\hbar
v_\mathrm{F}{\left(\mathbf{{m}}\times\mathbf{{k}}\right)}_z\right)}^2}.\nonumber
\end{eqnarray}

In the above, we have $(2\times 2)$ energy bands represented by two indexes,
{\it{i.e.}}, $s$ which has the same meaning as the spin index in Section
\ref{theory}, and $\tau $ which can be considered as the pseudo-spin index, which
refers to the mixing of the top or bottom surface states. By using the same
framework as in Section \ref{theory}, we can derive the torque field for each of
the energy band as:
\begin{equation} \label{GrindEQ__12_} 
{\mathbf{H}}^\mathrm{eff}_{s\tau
}=s\frac{J_\mathrm{ex}}{\sqrt{J_\mathrm{ex}^2+4\Delta^2}}\frac{n\hbar
v_\mathrm{F} }{\mu_0 M_s}\left(\hat{z}\times \mathbf{k}\right)
\end{equation} 
for the strong exchange limit ($J_\mathrm{ex}\gg \hbar v_\mathrm{F}
k_\mathrm{F}$), and
\begin{eqnarray} \label{GrindEQ__12a_} 
{\mathbf{H}}^\mathrm{eff}_{s\tau }&=& s
\frac{J_\mathrm{ex}}{\sqrt{4\hbar^2v_\mathrm{F}^2k_\mathrm{F}^2+4\Delta^2}}\frac{n\hbar
v_\mathrm{F} }{\mu_0 M_s}\left(\hat{z}\times \mathbf{k}\right)\nonumber\\
&+&s\tau \frac{\hbar v_\mathrm{F} J_\mathrm{ex}
(\mathbf{m}\times\mathbf{k})_z}{2(\hbar^2v_\mathrm{F}^2k_\mathrm{F}^2+\Delta^2)}\frac{n\hbar
v_\mathrm{F} }{\mu_0 M_s}\left(\hat{z}\times \mathbf{k}\right)
\end{eqnarray} 
for the weak exchange limit ($J_\mathrm{ex}\ll \hbar v_\mathrm{F}
k_\mathrm{F}$).

Let us first consider the strong exchange limit. If the Fermi energy $E_F=0$, only the two lower bands contribute to the torque
field in Eq. \eqref{GrindEQ__12_}, {\it{i.e.}}, $s=-1$ and $\tau =\pm 1$.
Interestingly, the torque field in Eq. \eqref{GrindEQ__12_} does not depend on
the band index $\tau $. Therefore, the total torque field is given by
${\mathbf{H}}^\mathrm{eff}=\sum_{s,\tau }{{\mathbf{H}}^\mathrm{eff}_{s\tau
}f_{s\tau }}$, with $f_{s\tau }$ being the distribution function for each energy
bands. The total torque field is given as
\begin{equation} \label{GrindEQ__13_} 
{\mathbf{H}}^\mathrm{eff}=\eta\left(\hat{z}\times
{\mathbf{j}}^{ss}_e\right),
\end{equation} 
where
\begin{equation}\label{eta}
\eta=\frac{\eta_0}{\sqrt{1+4{(\Delta/J_{\mathrm{ex}})}^2}}.
\end{equation}
We see that the direction of the torque field in this case is still the same as
in Eq. \eqref{GrindEQ__6_}, {\it{i.e.}}, in the transverse direction with
respect to the applied current. However, the amplitude is now modified by a
factor which is a function of the tunneling element $\mathrm{\Delta }$ as in Eq. \ref{eta}. Figs.
\ref{Fig2} (a), (b) depict the dependence of the torque efficiency  $|\mathcal{T}|/j^{ss}_e$ 
on the tunneling element and the TI thickness, respectively. As expected, for a given surface state current $j^{ss}$, the torque efficiency is
reduced when the $\mathrm{\Delta }$ increases, a trend which can also be
discerned from Eq. \eqref{GrindEQ__12_}.

One can check the underlying physics of the torque field
reduction by assuming the two surface states have the same helicity. By
replacing $-\mathcal{H}_\mathrm{TI}$ by $\mathcal{H}_\mathrm{TI}$ in the bottom
right block of Eq. (\ref{GrindEQ__9_}), one can find the torque field is
$\mathbf{H}^\mathrm{eff}=\eta_0\left(\hat{z}\times \mathbf{k}\right)$,
which is independent of the tunnel coupling $\Delta$. Therefore, we can conclude
that it is the coupling between surface states with opposite helicities which
reduces the net torque.

\subsection{Thickness optimization}

	\begin{figure}
		\includegraphics[width=0.5\textwidth]{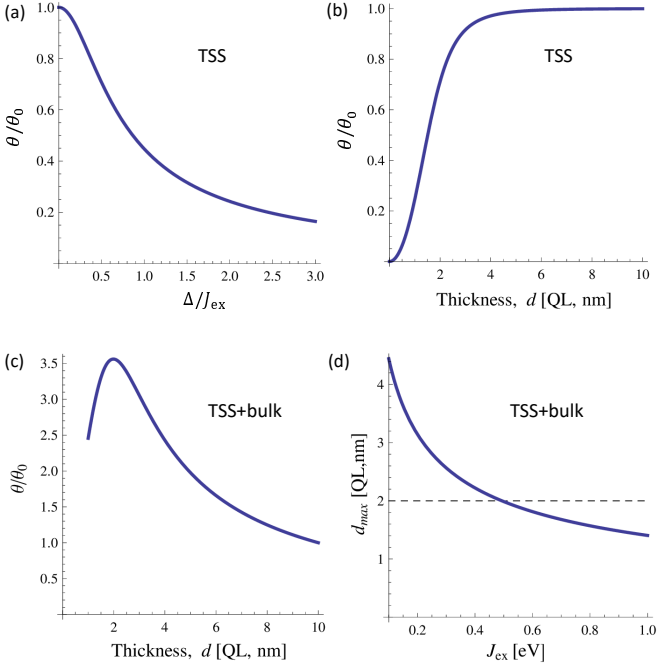}
\caption{(Color online) (a), (b) Spin torque efficiency induced by topological
 surface states (TSS) in an ideal TI in the absence of the bulk channel. 
(a) Spin torque as a function of the tunneling
coupling $\Delta$. (b) Spin torque as a function of thickness 
$d$, $J_{\mathrm{ex}}=0.1$ eV.  (c), (d) Spin torque efficiency
 induced by TSS in the presence of the bulk channel. (c)  Torque 
efficiency as a function of thickness. Comparing to the efficiency in
 thick film limit ($\theta_0$ normalized to 1), the
maximum efficiency is about 3.5 times larger at an optimal thickness
 $d_{max}\approx 3\
\mathrm{nm}$. Parameters used: $J_\mathrm{ex}=0.1\ \mathrm{eV}$, $G_\mathrm{TI}[e^2/h]\sim 19\ d$. (d) The value of
optimal thickness as a function of exchange coupling for various TI parameter
$B_1$. The horizontal dashed line represents the effective thickness of the TI
surface states ($\sim$ 2 nm). The peak of torque efficiency is only observed if the
value of $d_{max}$ is above the dashed line. ${\theta }_0$ is the efficiency at the 
limit of thick TI film. Parameters used: $B_1=0.1\ \mathrm{eV\ nm^2}$ \cite{Zhang:nat10}}\label{Fig2}
\end{figure}

As discussed in previous sections, decreasing TI thickness gives rise to two 
opposite effects on the spin torque: torque reduction due to the strong 
surface hybridization, and torque enhancement due to the weak bulk 
channel. In this section, we will thus derive the optimal thickness to achieve maximized spin torque.
 
As the thickness of TI film is reduced, the conductance of the TI film is also
diminished \cite{Ban:prl12}. In extremely thin TI films, where the thickness is
below 10 nm, the conductance is function of the thickness
$G_\mathrm{TI}\left(d\right)=a_G+b_Gd\ $ [in units of $e^2/h$]. Meanwhile, in
the thick-film limit, the conductance approaches an almost constant value
\cite{Ban:prl12}. Experimentally, $b_G\approx 19\
\mathrm{nm}^{-1}$ for $\mathrm{Bi_2Se_3}$ thin films \cite{Ban:prl12}.

 The spin torque efficiency, which is the ratio between spin torque and charge current
flowing in the TI, {\it{i.e.}}, $\theta= \mathcal{T}/j_{\mathrm{TI}}$, is then

\begin{equation}
\theta =\underbrace{\frac{\eta_0}{\sqrt{1+4({B_1\pi^2/d^2 J_{\mathrm
ex}})^2}}}_{f_1\left(d\right) increases\ with\ d}\
\underbrace{\left(\frac{G^{ss}_\mathrm{TI}}{a_G+b_Gd}\right)}_{f_2\left(d\right)
decreases\ with\ d}
\end{equation} 

The above expression is the product of two functions $f_1\left(d\right)\cdot
f_2\left(d\right)$, where $f_1$ ($f_2$) increases (decreases) with $d$. In the
limit of large $d$,  $f_1(d)\sim \eta_0$, and $f_2(d)$ is relatively constant with the
change in $d$ as the conductance of the TI film is almost unchanged \cite{Ban:prl12},
 which means that the torque efficiency at large TI thickness
would be almost constant. However, in the range of small $d$, the torque
efficiency is more strongly dependent on $d$. In Fig. \ref{Fig2} (c), the torque
efficiency first increases as the thickness is reduced. This is due to the
reduction of the bulk contribution to the TI conductance, which enhances the
spin-polarized current from the surface state channel. As the thickness is
further reduced, the torque efficiency becomes diminished as the hybridization
effect becomes the dominant factor. Thus, the torque efficiency reaches a
maximum peak at some optimal value of thickness $d_\mathrm{max}$.

Now we focus on the maximum efficiency and the corresponding thickness.
Basically, these two quantities depend on system's parameters such as $B_1,\
J_\mathrm{ex}$. For simplicity, we consider the case where
$G_\mathrm{TI}\left(d\right)=b_Gd$, for which the maximum efficiency can be
analytically found as
\begin{equation}
\theta_\mathrm{max}=\frac{\eta_0}{\sqrt{4{\pi }^2B_1/J_\mathrm{ex}\
}}\left(\frac{G^{ss}_\mathrm{TI}}{b_G}\right),
\end{equation}
corresponding to the optimal thickness of
\begin{equation}
d_\mathrm{max}=\sqrt{\frac{{2\pi }^2B_1}{J_\mathrm{ex}}}.
\end{equation}
Taking value of the tunneling parameter $B_1=0.1\ \mathrm{eV\ nm^2}$, and the
exchange coupling $J_\mathrm{ex}=0.1-0.5\ \mathrm{eV}$, the optimal thickness is
estimated to be $d_\mathrm{max}=2-5\ \mathrm{nm}$. We note here that, for the
maximum torque to be observed, the optimal thickness should not be below the
effective thickness of the TI surface states, {\it{i.e.}}, $d_\mathrm{max}>2\
\mathrm{nm}$ \cite{Ban:prl12,Lin:prb09} (see Fig. \ref{Fig2} (d)). 

Similarly, we can estimate the optimal thickness in the weak exchange limit,
 which is given by $d_\mathrm{max}=\sqrt{\pi^2B_1/\hbar v_\mathrm{F}k_\mathrm{F}}$. 
With the Fermi momentum $k_F = 0.07 - 0.14\  \AA^{-1}$ \cite{Hoe:pnas14,Ban:prl12},
 and the Fermi velocity $v_F\sim 5\times 10^5$ m/s \cite{Zhang:nat10}, the optimal 
thickness is estimated to be in the range of $\sim 1.4 - 2$ nm. This range is 
almost below the effective thickness of the TI surface state, and thus the
 optimal spin torque may not be observed in the weak exchange regime. 

\section{Conclusion}

In this paper, we have analytically derived the spin-orbit torque induced by a
thin TI film. First, we showed that in thin TI films, the spin torque is
reduced due to the surface hybridization effect. This decrease is a consequence
of the opposite helicities of the surface states on the two surfaces. On the
other hand, lowering the thickness of the TI film may have a positive effect on
the spin torque by reducing the bulk channel contribution to the total
conductance. This increases the proportion of current flowing through the
surface states, which constitutes the source of the current-induced torque on
the magnetic layer. Due to these opposing trends, there thus exists an optimal
TI thickness at which the torque efficiency is maximized. Based on typical
experimental parameter values, we found that the torque efficiency at the
optimal thickness to be several times larger than the torque values at the thick
TI-layer limit. Our prediction for the optimal thickness is within the practical
range which can be verified experimentally.

\begin{acknowledgments}
We acknowledge the financial support of MOE Tier II grant MOE2013-T2-2-125 (NUS
Grant No. R-263-000-B10-112), and the National Research Foundation of Singapore
under the CRP Programs “Next Generation Spin Torque Memories: From Fundamental
Physics to Applications” NRF-CRP12-2013-01 and “Non-Volatile Magnetic Logic and
Memory Integrated Circuit Devices” NRF-CRP9-2011-01.
\end{acknowledgments}


%

\end{document}